\documentclass[twocolumn,showpacs,preprintnumbers,amsmath,amssymb,aps,prl,
superscriptaddress
]{revtex4}
\usepackage{graphicx}
\usepackage{times}
\usepackage{color}
\usepackage[latin1]{inputenc}

\begin{document}

\title{Molecular superfluid phase in 
one-dimensional multicomponent fermionic cold atoms} 

\author{S. Capponi}
\email{capponi@irsamc.ups-tlse.fr} \affiliation{Laboratoire de Physique Th\'eorique, IRSAMC, CNRS, Universit\'e Paul Sabatier, 
 31062 Toulouse, France.}
\author{G. Roux} \affiliation{Laboratoire de Physique Th\'eorique, IRSAMC, CNRS, Universit\'e Paul Sabatier, 
 31062 Toulouse, France.}
\author{P. Lecheminant} \affiliation{Laboratoire de Physique Th\'eorique et Mod\'elisation,
Universit\'e de Cergy-Pontoise, CNRS, 2 Avenue Adolphe Chauvin, 95302
Cergy-Pontoise, France.}
\author{P. Azaria} \affiliation{Laboratoire de Physique Th\'eorique de la Matiere
Condens\'ee, Universit\'e Pierre et Marie Curie, CNRS, 4 Place
Jussieu, 75005 Paris, France.}
\author{E. Boulat} \affiliation{Laboratoire Mat\'eriaux et Ph\'enom\`enes Quantiques, Universit\'e Paris Diderot - Paris 7, CNRS,
75205 Paris Cedex 13, France.}
\author{S.~R. White} \affiliation{Department of Physics and Astronomy, University of California, 
Irvine, CA 92697, USA.}

\date{\today}
\pacs{
{03.75.Mn}, 
{71.10.Fd}, 
{03.75.Ss}, 
{71.10.Pm} 
}

\begin{abstract}
We study a simple model of $N$-component fermions with contact
interactions which describes fermionic atoms with $N=2F+1$ hyperfine
states loaded into a one-dimensional optical lattice.  We show by means
of analytical and numerical approaches that, for attractive
interaction, a quasi-long-range molecular superfluid phase emerges at
low density.  In such a phase, the pairing instability is strongly
suppressed and the leading instability is formed from bound-states
made of $N$ fermions. At small density, the molecular superfluid phase is generic and exists for a wide range of 
attractive contact interactions without an SU($N$) symmetry between the hyperfine states. 
\end{abstract}

\maketitle

Because of rapid progress in recent years, cold atom systems have become
a major field of research for investigating the physics of strong
correlations in a widely tunable range and in unprecedentedly clean
systems~\cite{review}.  Ultracold atomic systems also offer direct
access to the study of spin degeneracy since the hyperfine spin $F$
can be larger than 1/2, resulting in $2F+1$ hyperfine states.  In
non-magnetic traps, such as optical traps, this high degeneracy might give
rise to novel exotic quantum phases.  The superfluid state of
optically trapped alkali fermions with hyperfine spin $F > 1/2$ has
been studied with an emphasis on the general structure of the
large-spin Cooper pairs~\cite{ho}.  The spin-degeneracy in fermionic
atoms is also expected to give rise to more complex superfluid phases.
In particular, a molecular superfluid (MS) phase might be stabilized where
more than two fermions form a bound state.  Such a non-trivial
superfluid behavior has already been found in different contexts.  In
nuclear physics, a four-particle condensate---the $\alpha$ particle---is 
favored over deuteron condensation at low densities~\cite{schuck} and
it may have implications for light nuclei and asymmetric matter in
nuclear stars~\cite{deanlee}.  This quartet condensation can also
occur in  semiconductors with the formation of
biexcitons~\cite{nozieres}.  A quartetting phase, which stems from the
pairing of Cooper pairs, has also been found in a model of
one-dimensional (1D) Josephson junctions~\cite{doucot}. 
A similar phase has also been reported in exact-diagonalization calculations
of the two-dimensional t-J model at low doping \cite{poilblanc}. 
More recently, the emergence of quartets  and
triplets (three-fermion bound states) has been proposed to occur in
the context of ultracold fermionic atoms~\cite{Wu2005,phle,miyake,Rapp}. 

In view of this  increasing interest 
in the formation of complex superfluid 
condensates,  it would be highly desirable to have 
at one's disposal a  
simple paradigmatic $N$-component fermionic model 
which displays this exotic physics.
It will be shown in this letter that such a model 
is provided by the 1D $N$-component fermionic Hubbard model
with attractive contact interaction:
\begin{equation}
{\cal H}
= -t \sum_{i,\alpha} [c^{\dagger}_{\alpha,i} c_{\alpha,i+1} 
+ {\rm H.c.} ]+\frac{U}{2} \sum_i n_i^2,
\label{hubbardS}
\end{equation}
where $c^{\dagger}_{\alpha,i}$ is the fermion creation operator
corresponding to the $N=2F+1$ hyperfine states $\alpha = 1,\ldots,N$ and
$n_i = \sum_{\alpha} c^{\dagger}_{\alpha,i} c_{\alpha,i}$ is the
density at site $i$.  Model~(\ref{hubbardS})  displays an extended
U($N$)$=$U(1)$\times$SU($N$) symmetry 
and it has been recently introduced 
in the context of  ultracold fermionic atoms~\cite{honerkamp}. 
A possible experimental realization of this
model (\ref{hubbardS}) for $N=3$ would be a system of $^{6}$Li atoms loaded into a
1D optical lattice with a carefully tuned combination of external magnetic and 
optical fields to make three internal states exhibit SU(3)
symmetry~\cite{greiter}. The $N=4$ case might also be relevant to the
optical trap of four 
hyperfine states of $^{40}$K ($F=9/2$ atoms)~\cite{demarco}.
The SU($N$) symmetry of Eq. (\ref{hubbardS}) has an important 
consequence since, when $N>2$,  
even for $U<0$ there can be no {\it pairing} between fermions:
there is no way to form a SU($N$) singlet with only two fermions. The only
superfluid instability that can be stabilized is a molecular one where 
$N$ fermions 
form a SU($N$) singlet: $M_i^{\dagger} = c_{1,i}^{\dagger} c_{2,i}^{\dagger}
\cdots c_{N,i}^{\dagger}$. 
In this letter, we shall show, by means of a combination of analytical
and numerical results obtained by the density-matrix renormalization group
(DMRG) technique~\cite{dmrg}, that this MS phase emerges
in the phase diagram of  model (\ref{hubbardS}) for
$U<0$ and at {\it at small enough density $n$}.
The latter phase is not an artifact of the extended SU($N$) symmetry
of model (\ref{hubbardS})---it is robust
to symmetry breaking terms toward more realistic situations.
In this respect, we believe that the Hubbard model (\ref{hubbardS})
captures the main generic features responsible for the formation
of the MS  phase.

\textit{Low-energy approach.} The low-energy effective field theory
corresponding to the SU($N$) Hubbard chain (\ref{hubbardS}) can be
derived, as usual, from the linearization at the two Fermi points
($\pm k_F$) of the dispersion relation of free $N$-component
fermions~\cite{bookboso,giamarchi}. The derivation of the low-energy
Hamiltonian is straightforward (see for instance
Ref.~\onlinecite{assaraf} for details) and, away from half-filling, it
separates into two commuting charge and spin pieces: ${\cal H} = {\cal
H}_c + {\cal H}_s$.  This is the famous spin-charge separation which
is the hallmark of 1D electronic
systems~\cite{bookboso,giamarchi}.  Within this low-energy
description, the U(1) charge excitations are described by a free
massless bosonic field $\Phi$ with Hamiltonian density:
\begin{equation}
{\cal H}_c = \frac{v}{2} \left[\frac{1}{K}
\left(\partial_x \Phi \right)^2 + K
\left(\partial_x \Theta \right)^2 \right],
\label{luttbis}
\end{equation}
where $v$ and $K$ are respectively the charge velocity and the
Luttinger parameter. A perturbative estimate gives $v = v_F \left[1 +
U(N -1)/(\pi v_F)\right]^{1/2}$ and $K = \left[1 + U (N -1)/(\pi
v_F)\right]^{-1/2}$ with the Fermi velocity $v_F = 2t a \sin (k_F a)$
and lattice spacing $a$. For generic fillings, no umklapp terms appear
and the charge degrees of freedom display metallic properties in the
Luttinger liquid universality class~\cite{bookboso, giamarchi}. 

The hyperfine spin sector is described by the SU($N$)
Thirring model which is an integrable field theory~\cite{andrei}. 
For  attractive interaction ($U < 0$)
a spectral gap $m$ opens. The low-energy spectrum in the hyperfine spin sector 
consists of  $N-1$ branches with masses 
 $m_r = m \sin(\pi r /N)$  
($r = 1, \ldots, N-1$) \cite{andrei}. 
The dominant instability which governs the physics of this phase is the
one with the slowest decaying correlations at zero temperature.
Both the 1-particle Green function
$G(x)=\langle c^\dagger_{\alpha,i} c_{\alpha,i+x}\rangle$
and (onsite) 
pairing correlations $P(x)=\langle  c^\dagger_{\alpha,i} c^\dagger_{\beta,i}
c_{\beta,i+x} c_{\alpha,i+x}\rangle$ are short-range.
On the contrary, 
the equal-time density correlation $N(x) = \langle n_i n_{i+x} \rangle$
associated with a charge-density wave (CDW)
and the equal-time MS correlations $M(x) = \langle
M_i M_{i+x}^{\dagger}\rangle$ have the 
following power-law decay at long distance \cite{comment_eq}:
\begin{eqnarray}
N(x)& \sim& \cos(2 k_F x) x^{-2 K/N} \\
M(x) &\sim&  x^{-N/(2 K)} \quad \text{for }N\text{ even}\label{Neven.eq}\\
M(x) &\sim& \sin(k_F x) x^{-(K+N^2/K)/(2N)}\quad\text{for }N\text{ odd.}\label{Nodd.eq}
\end{eqnarray}

We thus see that CDW and MS instabilities 
compete and the key point of the analysis is the one
which dominates. At issue is  the value of the Luttinger parameter
$K$. In particular a
dominant MS instability requires   $K > N/2$ 
($K > N/\sqrt{3}$) for $N$ even (odd, respectively) 
and thus a fairly large value of $K$ which,
with only short range interaction, is not guaranteed. 
However a simple argument
suggests that this may be realized at sufficiently small density at large
negative $U$.  Indeed, when  
$n \ll 1$ and $\vert U\vert /t \gg 1$, a dilute gas of strongly bound $N$-fermion objects forms and
(\ref{hubbardS}) behaves as essentially \emph{free hardcore bosons} ($N$ even)
or {\it free fermions} ($N$ odd) 
with an effective hopping $t^N/\vert U \vert^{N-1}$.
One can  therefore  estimate  $M(x)$ in this limit
as the free bosonic Green function,
$M(x) \sim x^{-1/2}$~\cite{haldane,giamarchi}, 
when $N$ is even and, as the free fermion Green
function, $M(x) \sim \sin(k_Fx)/x$, when $N$ is odd. By comparing with
Eqs.~(\ref{Neven.eq}, \ref{Nodd.eq}), we  deduce an upper bound
for $K$ which is $K_{ \rm max} = N$ \cite{comment}. 
From the perturbative estimate 
we see that  $K > 1$ and $K$
increases with  $|U|$, so that
there is room to  stabilize an MS phase 
for  sufficiently strong attractive interaction and small density.  
In addition, at zero density, the $N$-component Fermi gas with
an SU($N$) symmetry is known to be exactly-solvable by means of the Bethe-ansatz approach
and bound states of $N$ fermions are formed for attractive interaction \cite{betheansatz}.
Outside these cases of infinite attractive interaction or vanishing density,
the existence and stability of this MS phase stem from the full
non-perturbative behavior of the Luttinger parameter $K$ as a function
of the density $n$ and the interaction $U$. We shall now
evaluate numerically this parameter in the simplest odd and even cases
$N=3,4$ by computing dominant correlations with the 
DMRG technique to conclude on the extension of the MS phase.
\begin{figure}[t]
\includegraphics[width=\columnwidth,clip]{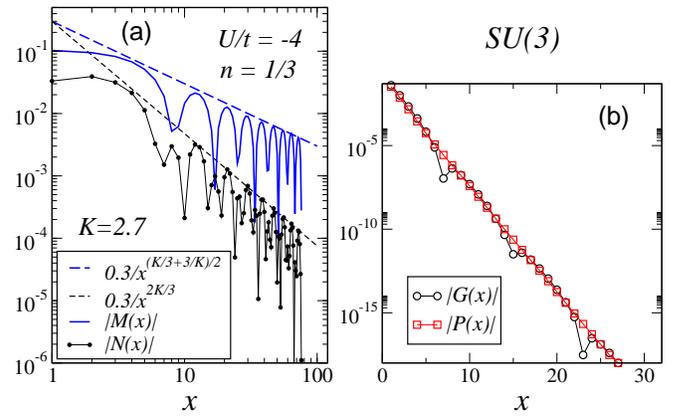}
\caption{(Color online) SU(3) model: triplet and density correlations
vs distance obtained by DMRG with $L=153$, $n=1/3$ and $U/t=-4$. (a) dominant
triplet over CDW correlations can both be fitted with $K=2.7$. We
also see the $k_F$ (respectively $2k_F$) oscillations of $M(x)$
(respectively $N(x)$). (b) 1-particle Green function $G(x)$ 
and pairing correlations $P(x)$ vs distance. Both are short-range 
and with the same correlation length $\xi=0.68$.}
\label{CorrSU3.fig}
\end{figure}
\begin{figure}[t]
\includegraphics[width=\columnwidth,clip]{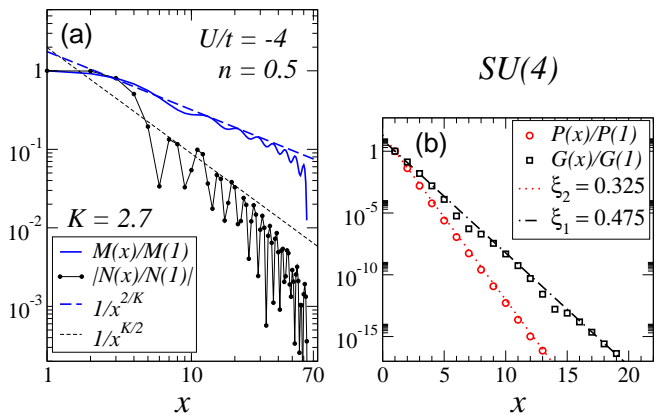}
\caption{(Color online) SU(4) model: (a) quartet and density
correlations vs distance obtained by DMRG with $L=128$ and $U/t=-4$ at
filling $n = 0.25$. The same $K \simeq 2.7$ is used to give a rough
estimate of the exponent. (b) 1- and 2-particle correlations 
vs distance. Both are short-range and the ratio of the
two correlation lengths is $\xi_2/\xi_1=0.68\simeq 1/\sqrt{2}$. }
\label{CorrSU4.fig}
\end{figure}
\begin{figure}[t]
\includegraphics[width=0.8\columnwidth,clip]{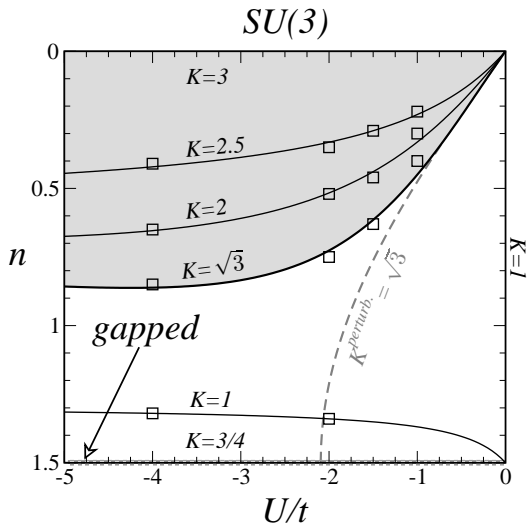}
\caption{SU(3) model: phase diagram showing the Luttinger parameter
$K$ vs filling $n$ and interaction $U$. The grey area is the
superfluid triplet phase. Lines are guide for the eyes which behavior satisfies the perturbative limit close to the point $(U/t=0,n=0)$.
We also plot the perturbative estimate separating both regions (see text),  
 that agrees with our numerical finding at small $|U|$ but deviates for larger values.}
\label{PhaseDiagramSU3.fig}
\end{figure}
\begin{figure}[t]
\includegraphics[width=0.8\columnwidth,clip]{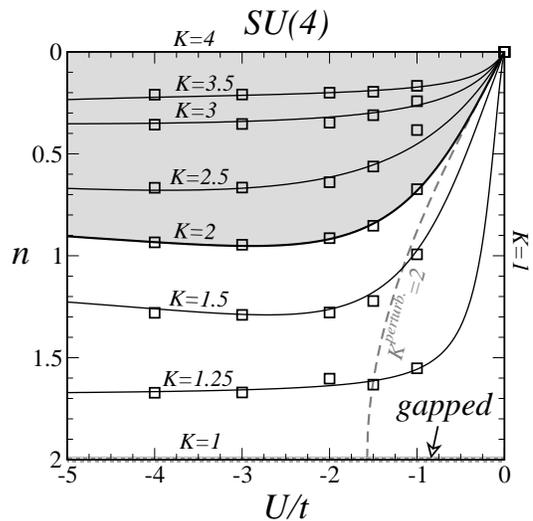}
\caption{
Same as Fig.~\ref{PhaseDiagramSU3.fig} for the SU(4) case. 
}
\label{PhaseDiagramSU4.fig}
\end{figure}

\textit{Numerical Results.} We have performed extensive DMRG
calculations
for both the $N=3$ and $N=4$ cases and for a wide range
of densities $n$ and interactions $U$ \cite{note,epaps}. 
We show in Fig.~\ref{CorrSU3.fig}(a-b) and
Fig.~\ref{CorrSU4.fig}(a-b) our data for 
$N=3$ and $N=4$ respectively at  typical values of  $n$
and $U= -4t$. 
In both cases, 
and in agreement with the
low-energy approach, we find that a gap opens in the hyperfine spin 
sector and that
the one-particle and pairing correlations are always short ranged. 
From  Fig.~\ref{CorrSU3.fig}(b) and
Fig.~\ref{CorrSU4.fig}(b), 
one can compute the one- and two-particle correlation lengths $\xi$ that is 
expected to vary as 
the inverse gap~: $\xi \sim 1/m$. We find that the ratio $R=m_1/m_2$
is close to $1$ and $1/\sqrt2$ respectively for $N=3$ and $N=4$ as expected 
from the low-energy approach. 
In contrast,
we see in Fig.~\ref{CorrSU3.fig}(a) and
Fig.~\ref{CorrSU4.fig}(a)  that the density and MS
correlations  $N(x)$ and $M(x)$
display power-law behavior. 
Clearly, triplet and quartet correlations dominate over CDW 
at these  densities ($n=1/3$ for  $N=3$ and  $n=0.5$ for  $N=4$). The phase
diagrams for both SU(3) and SU(4) models are presented in 
Fig.~\ref{PhaseDiagramSU3.fig} and Fig.~\ref{PhaseDiagramSU4.fig}
which give a map of $K$ vs interaction and density. 
The values of $K$ were obtained from the power-law behavior of the molecular
correlation $M(x)$ using Eqs.~(\ref{Neven.eq}, \ref{Nodd.eq}).  
We find that 
triplet and quartet superfluid phases emerge 
in a wide portion of the phase diagrams (grey area) 
separated from a CDW phase by a cross-over line $n_c(U)$.  
As the density decreases from half-filling ($n=N/2$) to 0, 
$K$ increases from $N/4$ to $N$ for
moderate or strong $|U|/t$, as 
expected from the strong-coupling argument.  
Interestingly enough,
the MS phase extends to small values of $U$ at sufficiently
small densities. We also observe that the curve $n_c(U)$ 
is likely to saturate
in the strong coupling limit~\cite{epaps}. 
For example in the SU(4) case, the $K(n)$ function is almost $U$-independent 
for $\vert U \vert /t > 2$ so that lines of equal $K$ are
parallel to the $U$ axis. 

\begin{figure}[t]
\includegraphics[width=\columnwidth,clip]{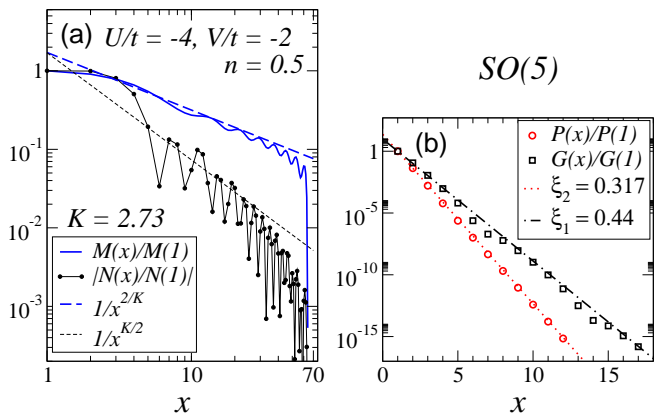}
\caption{SO(5) model~: (a) 
Quartet and density correlations for $U/t=-4$ and 
$V/t=-2$ at the density n=0.5. 
The value of the Luttinger parameter is $K=2.7$. 
(b) 1- and 2-particle correlations vs distance. 
Both are short-range and the ratio of the
two correlation lengths is $\xi_2/\xi_1=0.72\simeq 1/\sqrt{2}$. }
\label{CorrWithBCS_SO5.fig}
\end{figure}

\textit{Effect of symmetry breaking perturbations.} 
At this point, the natural question is whether the molecular superfluid
phases survives to the breaking of the SU($N$) symmetry.  
This is an important 
question since in most of the realistic situations, 
the actual symmetry is expected to be
much smaller. Part of the answer is given in 1D systems by the
accepted view that, at sufficiently low energies and for 
generic interactions, the dynamical symmetry is most
likely to be enlarged~\cite{balents}:
though the SU($N$) symmetry is not an exact symmetry, 
it is physically meaningful as an effective low-energy theory. 
As an example, we consider
the SU(4) case relevant for spin-3/2 cold atoms and add to 
the Hubbard Hamiltonian (\ref{hubbardS})  a singlet-pairing coupling 
$ V \sum_{i} 
P_{00,i}^{\dagger} P_{00,i}$ where 
$P_{00,i}^{\dagger} = c^{\dagger}_{3/2,i}c^{\dagger}_{-3/2,i} -
c^{\dagger}_{1/2,i}c^{\dagger}_{-1/2,i}$.
As shown
in Ref.~\onlinecite{zhang}, the pairing term reduces the SU(4) symmetry down 
to SO(5).  We  show typical data
for $U/t = -4$ and $V/t =-2$ at the density $n=1/2$ 
in Fig.~\ref{CorrWithBCS_SO5.fig}.
We clearly see that the equal-time pairing correlation function $P(x) = \langle 
P_{00,i}^{\dagger} P_{00,i+x}\rangle$ admits an exponential decay i.e. 
there is no BCS instability. In contrast, quartet correlations are (quasi) 
long ranged
and dominate over CDW ones. 
Remarkably, we observe from Fig. ~\ref{CorrWithBCS_SO5.fig}
that the gap ratio $R$ is very similar to the one for the full SU(4) symmetric
model when $V=0$. This means that the SU(4) model (\ref{hubbardS}) 
is a very good starting point to explore the main features
of the quartet phase.
Of course, for large negative $V$, 
a BCS phase does appear~\cite{Capponi2007} but the main point here
is to show that  the  quartet molecular phase is not an artifact
of the SU(4) symmetry and 
does exist in more realistic models~\cite{commentlutt}. 
A more detailed study will be presented elsewhere.

{\it Concluding remarks.} We have shown that a quasi-long-range
general MS phase can emerge in 1D for attractive
interactions at low density. This 1D phase, characterized by a bound-state made of $N$ fermions,
can be viewed as a nematic Luttinger liquid and 
a simple paradigmatic model to describe its main physical properties is
the attractive SU($N$) Hubbard chain~(\ref{hubbardS}).  The triplet and
quartet phases in the simplest $N=3,4$ cases might be explored
experimentally in the context of spinor ultracold fermionic atoms.  As
a first step, we have assumed here a homogeneous optical lattice and
neglect in the first approximation the parabolic confining potential
of the atomic trap.  We expect that this potential will not affect
the properties of this molecular phase at low density. Such an effect
could be investigated by DMRG calculations for
quantitative comparisons~\cite{kollath}. 
In the context of cold atoms experiments,
 the triplet and
quartet phases can be probed by radio-frequency spectroscopy
to measure the excitation gaps of the successive triplet-quartet
dissociation process. We hope that future experiments in ultracold
fermionic atoms will reveal the existence of these triplet and quartet
phases.

\acknowledgments 
We would like to thank F.~H.~L.~Essler, P.~Schuck,
G.~V.~Shlyapnikov, and A. M. Tsvelik for useful discussions
and their interest. 
SC and GR thank IDRIS
(Orsay, France) and CALMIP (Toulouse, France) for use of
supercomputers. SRW acknowledges the support of the NSF under
grant DMR-0605444.


\begin{thebibliography}{99}
\bibitem{comment_eq} The even-$N$ case was treated in Ref.~\onlinecite{phle}.
\bibitem{comment} One can show that the lower bound of K  
is $N/4$ which is reached in the vicinity of half-filling ($n=N/2$).
In addition, 
we expect that the
system is fully gapped as soon as $U < 0$ for $N > 2$ at half-filling.
\bibitem{commentlutt} 
In fact, the quartetting phase also emerges in a problem
with no extended SO(5) symmetry.
For instance, we can consider a model of two
spin-1/2 Hubbard chains coupled by an interchain density-density interaction.
We can show that a quartetting phase appears at low-density
for attractive interactions.
\end{thebibliography}
\end{document}